\documentclass[12pt]{iopart}
\usepackage{graphicx}

\begin{document}

\title[Thin-film-like BSCCO single crystals by mechanical exfoliation]{Thin-film-like high temperature superconducting $\mathrm{Bi_2Sr_2CaCu_2O_{8+}}{_x}$ single crystal by mechanical exfoliation}

\author{X Wang$^1$, L X You$^1$\footnote{Author to whom any correspondence should be addressed}, X M Xie$^1$, X Y Yang$^1$, C T Lin$^2$ and M H Jiang$^1$}

\address{$^1$ State Key Laboratory of Functional Materials for Informatics, Shanghai Institute of Microsystem and Information Technology (SIMIT), Chinese Academy of Science (CAS), Shanghai 200050, China}

\address{$^2$ Max-Planck-Institut f\"ur Festk\"orperforschung, Heisenbergstrasse 1, D-70569 Stuttgart, Germany}

\ead{lxyou@mail.sim.ac.cn}

\begin{abstract}
Thin-film-like high temperature superconducting $\mathrm{Bi_2Sr_2CaCu_2O_{8+}}{_x}$ single crystals (TSB) with a typical thickness of 20-200 nm is produced by mechanical exfoliation. The TSB, which shows excellent superconductivity, can adhere firmly to the substrate by van der Waals force. The TSB may find important applications in fabricating high temperature superconducting devices. Intrinsic Josephson junctions made of the TSB are demonstrated. A very interesting finding is that the TSB can follow exactly the topology of the substrate, making it possible to fabricate superconducting devices with artificial structures by using substrate template, for instance, step-edge Josephson junctions.
\end{abstract}


\submitto{\NJP}

\maketitle

\section{Introduction}

Superconducting thin films are essential for superconducting devices. An ideal thin film shall be defect free, like a high quality bulk single crystal, except that the thickness is much smaller. However, defects are unavoidable due to substrate mismatch, dislocations, grain boundaries etc. Up to now, the quality for thin films, especially for high temperature superconducting oxide thin films, is still the limiting factor for device performances and yield, etc.

For highly anisotropic bulk single crystals, it is possible to obtain ultra thin single crystals by using scotch tape, a method often referred to as mechanical exfoliation. With this method, Geim et al first reported a new material of graphene, a thin atomic layer of graphite\cite{Novoselov:Science04,Geim:Science09}. This material, formerly believed to be non-existing, has triggered a worldwide interest. The method was also successfully demonstrated in other quasi two dimensional materials including high temperature superconducting $\mathrm{Bi_2Sr_2CaCu_2O_{8+}}{_x}$ (HTS BSCCO)\cite{Novoselov:PNAS05}. However, the obtained single layer (single cell) BSCCO was reported to be insulating. The result is still intriguing, as it points out a way for the preparation of thin-film-like high quality superconducting BSCCO single  crystals, which will be of great importance not only for high quality devices and applications, but also for fundamental research such as two dimensional physics including superconductivity.

In this manuscript, we report \textbf{T}hin-film-like high temperature \textbf{S}uperconducting \textbf{B}SCCO single crystals (TSB) with the thickness of 20-200 nm produced by mechanical exfoliation. The TSB can adhere firmly to the substrate by the van der Waals force and works like a thin film on a substrate, showing excellent superconducting property consistent with that of the bulk materials. The intrinsic Josephson junctions (IJJs)\cite{Kleiner:PRB94} made of a TSB are demonstrated. Also reported is an interesting observation that the surface of the TSB can follow exactly the substrate topology, envisaging microstructure engineering of the TSB which may find potential applications such as in the fabrication of step-edge Josephson junctions.

\section{Experiments and Discussions}

\begin{figure}
\begin{center}
\includegraphics{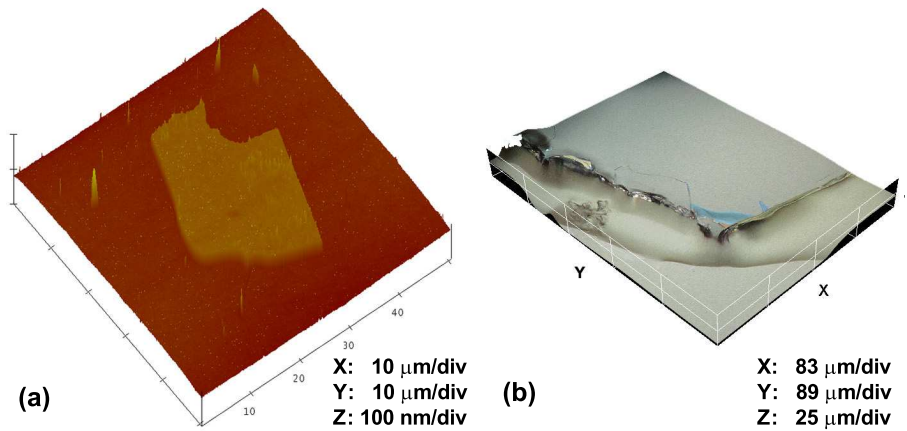}
\end{center}
\caption{(Color online) (a) AFM image of a TSB on a quartz substrate. The thickness of the TSB is 30 nm; (b) 3D image of a conventional BSCCO flake fixed on a substrate using polyimide.}
\label{image}
\end{figure}

BSCCO single crystals with a typical critical temperature $T_c \sim 85 $ K were grown using the traveling solvent floating zone method \cite{Lin:PhysicaC00}. The starting material is a BSCCO flake cleaved from a bulk BSCCO single crystal. The typical lateral size of the flake is about 1 mm by 1 mm and the thickness is a few tens of microns, which is often used for fabricating conventional IJJs. The flake is attached to a scotch tape or PE adhesive protective film, another scotch tape is then used to peel smaller flakes off the original BSCCO flake. By repeating this process for a few times, there will be many small visible and/or invisible BSCCO flakes attached to the two scotch tapes. Then we press one of the tapes to a substrate. Two kinds of substrates are used in experiments. One is a transparent quartz substrate; the other is a Si substrate with a capping layer of 300 nm thick SiO$_2$ which can give better optical contrast for easier TSB identification with an optical microscope. The SiO$_2$/Si is widely used as a substrate for mechanical exfoliated graphene. When the tape is removed from the substrate, some small invisible flakes become captured on the substrate surface, which can be registered by an optical microscope, however. The small flakes on the substrate may have different sizes and thickness. The interesting ones for electronic applications are those with the thickness from a few tens of nanometers up to about 200 nm, similar to the typical thickness of thin films. They often have the lateral size from a few microns to about a hundred microns. \Fref{image}(a) shows an atomic force microscopic image of a TSB on a quartz substrate, which has a lateral size of about 20 microns and the thickness is measured to be 30 nm (about 10 unit cells in the $c$-axis). In a sense, a TSB on a substrate can be regarded as a thin film grown on a substrate, then all the micro/nano-fabrication process in thin film technology can be adopted. Advantages of the TSB include its high quality, its compatibility with any kinds of substrates, and low interfacial stress due to van der Waals interaction between TSB and the substrate\cite{Novoselov:Science04,Li:JPD10}. It is worthy to note that the van der Waals force between the TSB and the substrate is strong enough so that the TSB sticks firmly to the substrate and does not drop off under moderate ultrasonic cleaning.

\begin{figure}
\begin{center}
\includegraphics{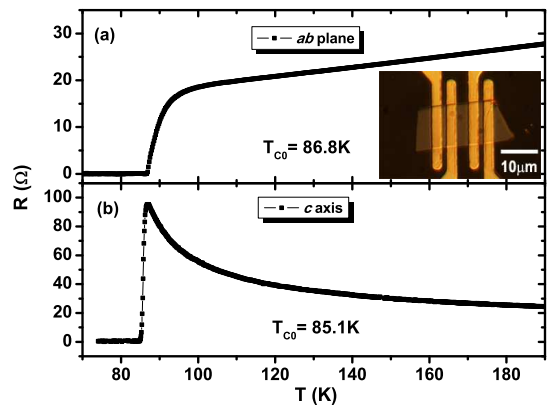}
\end{center}
\caption{(Color online) (a) R-T relation of a TSB in the $ab$ plane; (b) R-T relation of a TSB in the $c$-axis measured from a IJJs stack}
\label{R-T}
\end{figure}

As a strong anisotropic superconducting material, the superconducting properties of the TSB in the $ab$ plane and along the $c$-axis are crucial for the applications. For probing the $ab$ plane superconductivity, four gold electrodes in parallel with each other are evaporated on the surface of the TSB (see the inset of \fref{R-T}(a)). Since the $ab$ plane is parallel to the substrate, the resistance-temperature (R-T) relation in the $ab$ plane can be measured by the popular four-probe measurement. \Fref{R-T}(a) shows the $ab$ plane resistance-temperature (R-T) relation of a TSB with the thickness of 45 nm, which shows excellent superconductivity with a critical temperature $T_{C0//ab}$ of 86.8K. The $c$-axis R-T relation of another TSB sample is measured and shown in \fref{R-T}(b). The method of the measurement is different from the one for the $ab$ plane measurement and will be described later in this manuscript. The $c$-axis critical temperature $T_{C0\bot ab}$ is measured to be 85.1 K. The small difference of $T_C$ between $T_{C0//ab}$ and $T_{C0\bot ab}$ is due to the sample-to-sample variation.

The TSB has shown excellent superconductivity which has potential applications for fabricating superconducting devices. It is well-known that epitaxial HTS thin films grown by reactive magnetic sputtering or pulsed laser deposition may have very good uniformity along the $c$-axis. However,
the $ab$ plane often shows granular characteristics with grain boundaries, which is unexpected for many applications. As the TSB is directly separated from the high quality bulk single crystals, so the devices made of TSB should have better performance than those made of epitaxial thin films. Actually, thick BSCCO single crystal flakes with large sizes have been applied into fabricating IJJs. \Fref{image}(b) shows a conventional thick BSCCO flake glued on a substrate ready for IJJs' fabrication. Only a corner part of the sample is imaged since it is too `large' in comparison to the MLB. The lower plane is the surface of the substrate and the upper plane is the BSCCO flake. The middle part is the glue. The height between the surface of the BSCCO flake and the substrate is about 50 microns.

There are a few disadvantages of IJJs made of the conventional BSCCO single crystal flake. First, for the application as a detector/receiver, RF signal is often irradiated from a lens attached to the backside of the substrate to obtain a good coupling efficiency. BSCCO has a magnetic field penetration length of $\sim$ 100 $\mu$m along the $c$-axis\cite{Trunin:JC01}, which is close to the thickness of the BSCCO flake. The large and thick superconducting pedestal will absorb/block most of the signal, which is unfavorable in the detection efficiency. Secondly, The thickness of the BSCCO flake is often 4 orders of magnitude larger than the height of an IJJ, since the thickness of one IJJ is 1.5 nm and a typical stack of IJJs may include a few to a few tens of IJJs. From the perspective of thermal diffusion, the thick BSCCO pedestal blocks the heat diffusion from IJJs to the substrate since BSCCO has a large thermal resistance\cite{Verreet:SUST07}. On the other hand, the glue has an inferior thermal conductivity which turns to be another weak thermal diffusion part between IJJs and the substrate. Besides, it is also very hard to control the thickness and the levelness of the BSCCO flake by this gluing technology, which gives rise to the difficulty in precise alignment and a fine device structure achieved by the conventional photolithography or E-beam lithography. However, the thickness of TSB is much smaller in comparison to the magnetic penetration length, and there is no thick BSCCO pedestal and no glue for fixing the TSB to the substrate. As a result, those disadvantages can be effectively avoided or degraded if TSB is used to replace the conventional BSCCO flake for fabricating IJJs.

\begin{figure}
\begin{center}
\includegraphics{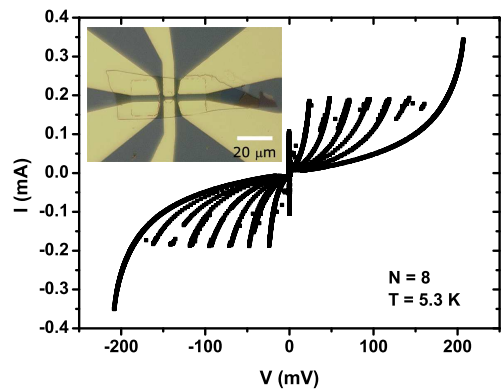}
\end{center}
\caption{(Color online) I-V curves of IJJs made of the TSB. The inset is an optical image of the sample}
\label{I-V}

\end{figure}

As a demonstration, we successfully fabricated IJJs using a TSB on a quartz substrate. The fabrication process is similar to the conventional IJJs' fabrication process in previous publications\cite{You:PRB05}. The inset of \fref{I-V} shows an optical image of a sample. The TSB is about 25 microns wide and more than 60 microns long, and the thickness of the TSB is measured to be $\sim$ 160 nm.

The R-T relation of the IJJs made of the TSB was measured and shown in \fref{R-T}(b). On one hand, it indicates the superconducting property of the IJJs; on the other hand, it reflects the superconductivity of the TSB along the $c$-axis. \Fref{I-V} shows the current-voltage (I-V) curves measured at 5.3 K. Eight quasi-particle branches have been traced by sweeping the bias current, which shows that the junction stack includes 8 junctions. These junctions have the consistent critical current of $\sim$ 0.2 mA, corresponding to the critical current density $j_c=$0.8 KA/cm$^2$. The critical current at the zero-voltage is slightly lower, which was often observed in IJJs fabricated on the conventional BSCCO flake. A possible explanation is that the superconductivity of the top junction might be degraded during the ion milling.

The thickness of the TSB is somehow random and the typical value is 20-200 nm. A thinner TSB down to one-unit-cell thickness can be produced, however, with elaborate mechanical exfoliation. Previous study showed that, with precise low energy Ar-ion milling, the thickness of a TSB can be further decreased and precisely controlled\cite{You:PRB05}. Better thickness conformity can be made by ion-mill thinning of thick TSBs. A ultrathin BSCCO single crystal with the thickness of 12 nm was demonstrated to have the same superconductivity with the bulk single crystal\cite{You:JAP05}. As a result, the TSB may find potential applications in fabricating devices, which were often made of ultra-thin films. Typical examples include HTS nanowire single photon detectors and hot electron bolometric mixers. It should be pointed out that the thickness of a TSB can go further down to a few nanometers, approaching its 2-dimension limit. Superconductivity can be suppressed by the superconductor-insulator transition\cite{You:JAP05}. Effect of thickness on the superconductivity of the TSB down to one unit cell is an interesting question worthy of further exploration.

\begin{figure}
\begin{center}
\includegraphics[width=16 cm]{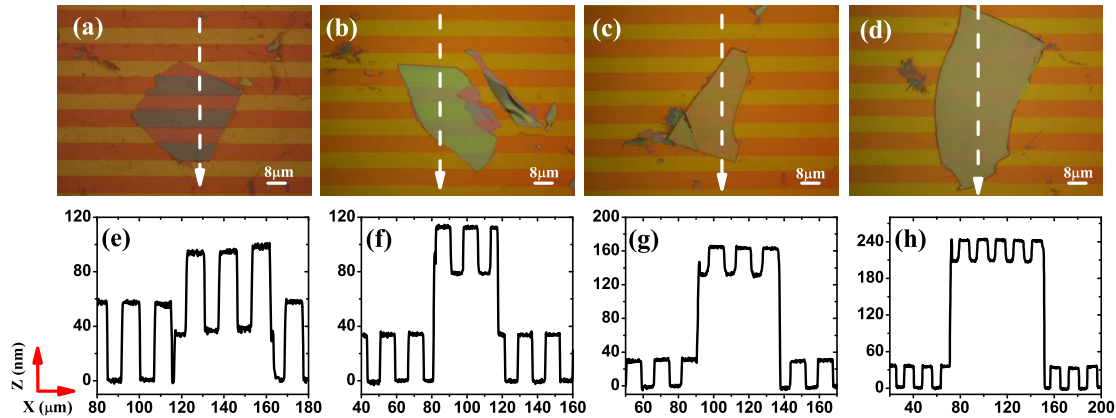}
\end{center}
\caption{(Color online) (a),(b),(c),(d) Four TSB samples on SiO$_2$/Si substrates with step structures; (e),(f),(g),(h) Scan lines across the TSB in (a),(b),(c),(d) respectively measured by a profilometer. In (a), the depth of the steps is 56 nm; in (b),(c)and (d), the depths of the steps are 33 nm. The thickness of the TSB in four images are 28nm, 77 nm, 133 nm and 208 nm respectively. The steps on the TSB were measured to have the same depth with the step on substrate.}
\label{stepTSB}
\end{figure}

One interesting property we observed is that the TSB can follow the topography of the substrate due to the strong TSB-substrate interaction. To make a better demonstration, SiO$_2$/Si substrates are prepared with a periodic step structure on the surface, which can be done by photolithography and Ar-ion milling. The period of steps is set to be 8 microns to make sure that a TSB can easily cover at least one step. Then the TSB is prepared on the step substrates. \Fref{stepTSB} shows the results of four samples. A 20 nm-thick gold layer is evaporated to protect the surface of TSB. It is interesting to notice that the TSB copies exactly the step structures on the substrate for various step heights and TSB thickness up to 200 nm. This result indicates that TSB is so `soft' that it can be regulated by morphology of the substrate due to the van der Waals interaction between the substrate and the TSB. Such substrate regulation was also experimentally observed for graphene and a few-layer graphene, and interpreted theoretically based on minimum energy analysis\cite{Li:JPD10,Stoberl:APL08}.

Since a TSB can follow the surface morphology of a substrate when there is a step on the substrate, it is possible to engineer the surface structure of TSB by using substrate templates. This gives us a new idea of fabricating step-edge Josephson junctions using single crystals instead of the popular thin film technology\cite{Hilgenkamp:RMP02}. Experimental justification of the TSB step-edge junction is still undergoing.

\section{Conclusions}

As a conclusion, the TSB with the thickness of 20-200 nm was obtained from BSCCO bulk single crystals by mechanical exfoliation, which can be transferred to a substrate. Due to the strong van der Waals interaction between the TSB and the substrate, the TSB adheres to the substrate, eliminating the need of any glue. TSB can follow exactly the substrate topology, creating an interesting possibility of structure engineering by using substrate template. The TSB has excellent superconductivity consistent with the bulk single crystal, which has potential applications in fabricating HTS devices, such as IJJs, single crystal step-edge Josephson junctions, nanowire single photon detectors and hot electron bolometric mixers. IJJs made of the TSB are successfully demonstrated. The TSB, when working together with Ar-ion milling, may also provide a good experimental platform to study low dimensional physics including superconductor-insulator transition.

\ack
This work is financed by Science and Technology Commission of Shanghai Municipality (Grant Nos. 08PJ1411200 and 08dz1400702), the National Natural Science Foundation of China (Grant No. 60801046), 973 program (Grant No. 2009CB929602) and Knowledge Innovation Program of CAS.

\end{document}